\documentclass[aps,prb,reprint,showpacs,groupedaddress]{revtex4-1}

\usepackage{graphicx}
\usepackage{graphics}
\usepackage{color}

\newcommand{\functderiv}[2]{\frac{\delta {#1}}{\delta {#2}}}
\newcommand{\del}[0]{\partial}
\newcommand{\trlog}[1]{\text{tr log}\, {#1}}
\newcommand{\curl}[1]{\nabla \times {#1}}

\newcommand{\tr}[1]{\text{tr }{#1}}

\begin{document}


\title{Quantum dynamics of local phase differences between reservoirs of driven interacting bosons separated by simple aperture arrays}


\author{T.J. Volkoff}
\author{K.B. Whaley}
\affiliation{Berkeley Quantum Information and Computation Center, UC Berkeley, Berkeley, CA 94720, U.S.A. \\ Department of Chemistry, UC Berkeley, Berkeley, CA 94720, U.S.A.}


\begin{abstract}
We present a derivation of the effective action for the relative phase of driven, aperture-coupled reservoirs of weakly-interacting condensed bosons from a (3+1)-D microscopic model with local U(1) gauge symmetry. We show that inclusion of local chemical potential and driving velocity fields as a gauge field allows derivation of the hydrodynamic equations of motion for the driven macroscopic phase differences across simple aperture arrays. For a single aperture, the current-phase equation for driven flow contains sinusoidal, linear, and current-bias contributions.  We compute the renormalization group (RG) beta function of the periodic potential in the effective action for small tunneling amplitudes and use this to analyze the temperature dependence of the low-energy current-phase relation, with application to the transition from linear to sinusoidal current-phase behavior observed in experiments by Hoskinson et al. \cite{packard} for liquid $^{4}$He driven through nanoaperture arrays.  Extension of the microscopic theory to a two-aperture array shows that interference between the microscopic tunneling contributions for individual apertures leads to an effective coupling between apertures which amplifies the Josephson oscillations in the array. The resulting multi-aperture current-phase equations are found to be equivalent to a set of equations for coupled pendula, with microscopically derived couplings. 
\end{abstract}


\maketitle

\section{\label{sec:intro}Introduction}
\subsection{Background}

The observation of coherent Josephson oscillations between driven reservoirs of liquid $^{4}$He separated by an array of nanometer-sized apertures\cite{packard} has stimulated research into thermal/quantum fluctuations of the macroscopic phase in aoerture array geometries\cite{pekker,chui}. These studies provide thermodynamic justification for the transition, observed in Ref.[\onlinecite{packard}], between two different current-phase relationships as a function of temperature below the lambda point: the linear regime (occurring at low temperatures, $T_{\lambda} -T > 5$mK ) in which the current depends linearly on the phase, and the ``weak-link" Josephson flow regime  in which the current has sinusoidal phase dependence (occurring for $T_{\lambda}-T < 0.8$mK). The linear current-phase relationship at low temperatures is thought to be due to independent phase slips occurring at individual apertures in the array in response to external driving by the hydrodynamic resonator.  As the temperature is increased toward the lambda point, the coherence length increases and the phase differences across individual apertures appear to become synchronized. It has been proposed that this results in coherent dissipative events, i.e. ``phase-slip avalanches," giving way eventually to coherent Josephson flow and a characteristic sinusoidal current-phase relation~\cite{packard,pekker,chui}.

No microscopic quantum mechanical explanation currently exists for this phenomenon. In order to justify the observed synchronization of the phase differences, Pekker, et al. 
postulate an effective long-range interaction between local phase gradients \cite{pekker} while Chui, et al. exploit the analogy between a Josephson junction array and classical coupled pendula to explore thermal phase fluctuations in an aperture array 
\cite{chui}. In this work, we derive an effective theory and equations of motion for the phase difference across a single aperture and a simple two-aperture array,
starting from a local U(1) gauge theory coupled to bosonic matter. The gauge field is necessitated by the presence of an external driving velocity which induces a ``vector potential" $v(r,t)$ and concomitant local chemical potential $\phi(r,t)$, analogous to the electromagnetic gauge field $A_{\mu}$ in the theory of Bardeen, Cooper and Schrieffer (BCS) for superconductivity.

Since our aim is a microscopic derivation of the equations of motion for the macroscopic phase differences across aperture arrays (Section \ref{sec:effective}) and an examination of the low-energy properties of the resulting current-phase relation across an aperture (Section \ref{sec:rg}), we employ here functional integral techniques rather than well known mean field or hydrodynamic techniques for bosonic systems (e.g., a gauged Gross-Pitaevskii (GP) equation\cite{hohenberg} for weakly-interacting Bose gases or a gauged two-fluid model~\cite{khalatnikov2} for $^{4}$He near the lambda point). The functional integral approach allows the equations of motion to be derived from the microscopic Lagrangian, as was demonstrated for the analogous case of superconducting systems by Ambegaokar et al. in Ref.~\cite{ambegaokar,ambegaokar2} (henceforth referred to as AES). Both the
stationary phase analyses and the perturbative renormalization group procedure in this work are most convenient to carry out using this formalism.

The fundamental variable of our effective theory is a gauge-invariant phase difference across an aperture: $\Delta \gamma (t) = \Delta \theta(t) +m\int \, dr \cdot v(r,t)$, where the integral is taken on a short line segment through the aperture. It contains contributions from the background phase texture $\nabla \theta(r,\tau)$ (the irrotational superfluid velocity) and the external driving velocity (the gauge field). We show that the action governing the gauge-invariant phase differences for simple aperture arrays provides a qualitative explanation for the experimental observations of $^4$He flow through nanoaperture arrays over a range of temperatures below $T_{\lambda}$~\cite{packard}.

As noted above, a related microscopic derivation for a superconducting system appears in AES, in which an effective theory is derived for the dynamics of a superconducting tunnel junction in terms of the macroscopic phase difference across the junction or the magnetic flux threading a superconducting quantum interference device.  Like AES, we shall be concerned here only with the dynamics of the low-energy degree of freedom in the system, namely, the macroscopic phase difference between junction-coupled Bose gases. In the present work, we focus on incorporating an externally imposed driving velocity into a gauge-invariant description of coupled reservoirs of weakly-interacting bosons, on determining the current-phase relation for this system in different parameter regimes dependent on the energy scale, and on using the results of this analysis to interpret the experimental observations of Ref.~\cite{packard}. We shall not undertake the further analysis of real-time current correlations, dissipation due to quasiparticles, or the effects of noise in the junction that was also made in AES. Explicit comparison between our results for driven, weakly-interacting bosons with the results of AES for superconducting systems will be given where relevant in the subsequent sections.
\subsection{Summary of results}
The microscopic analysis presented in this work shows that the main features of the transition from linear-to-sinusoidal Josephson flow as a function of temperature are apparent already in the one and two-aperture cases. Starting from a local U(1) gauge-invariant Lagrangian, we derive the effective action for one and two-aperture arrays. We first show that a perturbative expansion of the gauge theory can be used to derive the quantum hydrodynamical equations of motion for the driven superfluid. In particular, we show that the Josephson-Anderson equation for phase evolution \cite{packardJA} in gauge-invariant form, the circulation (superfluid fluxoid) quantization in the presence of a driving velocity field, and the London equation leading to the Hess-Fairbank effect \cite{hess} can all be derived from the stationary-phase approximation to the effective action.  
A Legendre transformation of the Euclidean effective action is then used to derive the current-phase relations for one and two-aperture arrays.  We show that for a single aperture, the current-phase relation is consistent with a potential composed of sinusoidal, linear, and quadratic terms, while for a two-aperture array we find that interference between the microscopic tunneling contributions for individual apertures leads to a coupling of the current-phase equations of the two-aperture system.

For the single aperture case, we then employ a weak-coupling renormalization group calculation to demonstrate the existence of temperature intervals in which the current-phase relation has predominantly linear or predominantly Josephson (sinusoidal) behavior. The critical temperatures separating these regions of linear and sinusoidal behavior are determined by relating the ratio of two coefficients in the rescaled effective action, each of which we calculate microscopically to one-loop order in perturbation theory, to the finite-temperature healing length.  Application of the theory to the experiment in Ref.~\cite{packard} on driven $^{4}$He flow through arrays of nanometer-sized apertures provides a rationalization for the transition between linear and sinusoidal current-phase relationships that was observed as the temperature was increased toward the lambda point.

\subsection{Outline}
In Section \ref{sec:model} we discuss the local U(1) gauge invariant Euclidean action used in the coherent-state functional integral and transform this action into a bilinear form in the real density field which can be analyzed using perturbation theory. In Section \ref{sec:effective} integration over the density field is performed and the resulting perturbation series for the full inverse Green's function is used to determine the effective action for the phase difference across a single aperture. We show that the stationary phase approximation to the perturbed action allows gauge invariant forms of several superfluid hydrodynamical equations to be derived, \textit{e.g.} the Josephson-Anderson equation for phase-difference evolution, the London equation for the gauge-invariant velocity, and circulation (superfluid fluxoid) quantization. The central result of this paper is the derivation of current-phase relationships for the single aperture and two-aperture array in Section \ref{sec:josephson}. We analyze the temperature dependence of the current-phase relation for a single aperture in the limit of small tunneling amplitude by computing the RG beta function of the coupling constant $E_{J}$ of the periodic potential and use this to analyze the experimental measurements of driven $^{4}$He flow through arrays of nanometer-sized apertures.  We summarize in Section \ref{sec:conclusion} and discuss potential directions for future research.

\section{\label{sec:model}The model}
We seek an effective theory for condensed, driven, weakly-interacting bosons separated by an array of one or two apertures in terms of local phase differences across the apertures. While our model includes only a local two-body potential, we will show that the main features of recent experimental results for liquid $^{4}$He flow through nanoaperture arrays~\cite{packard} are nevertheless already explained by the current analysis. Our starting point is the Hamiltonian in Eq.~(\ref{eqn:hamiltonian}) for the weakly-interacting Bose gas that is minimally coupled to a local chemical potential field $\phi(r,\tau)$ and a vector field $v(r,\tau)$ which will be interpreted as an external driving velocity. The Hamiltonian (without a tunneling term) is \begin{eqnarray} H[\hat{\psi}^{\dagger},\hat{\psi}] &=& \frac{1}{2m}\int \,  d^{3}r\bar{D}\hat{\psi}^{\dagger}(r)D\hat{\psi}(r)   +\frac{V_{0}}{2}\int d^{3}r \, \hat{\psi}^{\dagger \, 2}(r)\hat{\psi}^{2}(r) \nonumber \\  &+& m\int d^{3}r \, \phi(r)\hat{\psi}^{\dagger}(r)\hat{\psi}(r) + H_{\mathrm{ext}}[v,\phi] \label{eqn:hamiltonian}\end{eqnarray} The weak interaction is given by the usual delta function two-body potential, with strength $V_{0}$ (proportional to the $s$-wave scattering length), and $D=\nabla +imv(r,\tau)$ is the covariant derivative. $H_{\mathrm{ext}}[v,\phi]$ is a classical energy analogous to electromagnetic field energy in superconductors and depends only on external fields.

We then construct the coherent state path-integral Lagrangian, Eq.~(\ref{eqn:lagrangian}), corresponding to this Hamiltonian and additionally incorporate a single aperture tunneling term $T_{r,r'}$ that couples points $r$ and $r'$ on different sides of the aperture.  In the bosonic coherent state path integral, the Lagrangian is given by ($\hbar = k_{\mathrm{B}} = 1$): \begin{widetext} \begin{eqnarray} L[\psi, \psi^{*},\Delta ,v,\phi ] &=& \int d^{3}r \psi^{*}(r,\tau)(\partial_{\tau}+m\phi(r,\tau)-\mu)\psi (r,\tau) +\frac{1}{2m} \int d^{3}r \bar{D} \psi^{*}(r,\tau) D \psi(r,\tau ) \nonumber \\ &+& \int d^{3}rd^{3}r' \psi^{*}(r,\tau )T_{r,r'}\psi (r',\tau ) +\frac{V_{0}}{2} \int d^{3}r   \Delta^{*}(r,\tau)\Delta(r,\tau)  \nonumber \\ &-& \frac{V_{0}}{2} \int d^{3}r  \left[\Delta(r,\tau)\psi^{*}(r,\tau)\psi(r,\tau) - \Delta^{*}(r,\tau)\psi^{*}(r,\tau)\psi(r,\tau)\right] \nonumber \\ &+& \frac{mL^{2}}{2}\int d^{3}r (\nabla \times v(r,\tau))^{2} +\frac{mL^{2}}{2}\int d^{3}r (i\partial_{\tau}v(r,\tau)-\nabla \phi(r,\tau))^{2}  \label{eqn:lagrangian} \end{eqnarray} \end{widetext} 

Here, $\Delta (r,\tau)$ and $\Delta^{*}(r,\tau)$ are Hubbard-Stratonovich fields introduced to decouple the quartic interaction in the weakly interacting Bose gas,  $L$ has dimension of length, and $\tau$ is the imaginary time. In the grand canonical partition function, $Z(\mu,\beta)=\int e^{-\int_{0}^{\beta}d\tau L}$, the functional integration is over the fields $\psi$, $\psi^{*}$, $\Delta$, and $\Delta^{*}$ and also the gauge field (with the measure defined in the discretized expression for the coherent state path integral~\cite{negele}). The last two terms are derived from $H_{\mathrm{ext}}[v,\phi]$ in Eq.(\ref{eqn:hamiltonian}) and are analogous to the electromagnetic field energy in superconductors; the vorticity (circulation energy density) corresponding to the magnetic field energy density and an ``electric" energy density analogous to the electric field energy density. The fields $v(r,\tau)$ and $\phi(r,\tau)$ are analogues of the magnetic vector potential and local voltage of electrodynamics. These will be shown to satisfy stationary phase equations (Section \ref{sec:effective}) and we do not analyze fluctuations of the gauge field configurations.
 
If the tunneling matrix is multiplied by a U(1) parallel transporter via: \begin{equation} T_{r,r'} \rightarrow T_{r,r'}e^{im\int_{r}^{r'}dr \cdot v(r,\tau)} \end{equation} this Lagrangian is clearly invariant under $\psi(r,\tau) \rightarrow \psi(r,\tau)e^{i\Lambda (r,\tau)}$ (where $\Lambda(r,\tau)$ is real) as long as $v(r,\tau) \rightarrow v(r,\tau)-{1 \over m}\nabla \Lambda(r,\tau)$ and $\phi(r,\tau) \rightarrow \phi(r,\tau) - {i\over m}\partial_{\tau}\Lambda(r,\tau)$. Put another way, we are analyzing a local U(1) gauge theory for the superfluid where $(\phi(r,\tau),v(r,\tau))$ is the $\mathfrak{u}(1)$ gauge field. The gauge transformation of the 0-component is due to working in imaginary-time (i.e. the base-space for the U(1) principal bundle is a Euclidean manifold). In the analysis to follow, it will lead to e.g. an imaginary Josephson-Anderson equation, which must be Wick rotated to obtain the real-time equation. The mean-field equations of the gauged weakly-interacting Bose gas are the stationary phase equations of this bare action: $\functderiv{L}{\psi^{*}(r,\tau)}=0$ gives a gauged Gross-Pitaevskii equation \cite{hohenberg} for $\psi$, while $\functderiv{L}{\Delta^{*}(r,\tau)}=0 \Rightarrow  \Delta(r,\tau)  = -\vert \psi(r,\tau) \vert^{2}$ and $\functderiv{L}{\Delta(r,\tau)}=0 \Rightarrow  \Delta^{*}(r,\tau)  = \vert \psi(r,\tau) \vert^{2}$. Note that the Hubbard-Stratonovich fields are not complex conjugates! This is a peculiarity of the bosonic Hubbard-Stratonovich transformation. Since the action does not depend on space-time derivatives of $\Delta(r,\tau)$ or $\Delta^{*}(r,\tau)$, they may be taken as real constants at mean-field level.  In the following, we choose $\Delta(r,\tau) = -\Delta$ and $\Delta^{*}(r,\tau)=\Delta$ with $\Delta$ a real constant.

To isolate a local phase field, a polar decomposition can be made on $\psi$ and $\psi^{*}$, e.g. $\psi \rightarrow \sqrt{\rho(r,\tau)}e^{i\theta(r,\tau)}$. This transformation does not change the measure in the functional integral for the partition function. Physically it means we are considering a restricted ensemble, i.e. we consider only a single condensed mode in the path integral. This is our only explicit use of Bose-Einstein condensation of the weakly-interacting Bose gas in this work. The action corresponding to the resulting Lagrangian can be brought into bilinear form: 
\begin{widetext}\begin{eqnarray}
 S &=& \int^{\beta}_{0}d\tau 'd\tau \int d^{3}r'd^{3}r\bigg( \sqrt{\rho(r,\tau)}G^{-1}(r,\tau ; r', \tau ')\sqrt{\rho(r',\tau ')}+\frac{V_{0}}{2} \Delta^{2}\delta(r-r')\delta(\tau-\tau ') + \frac{mL^{2}}{2}(\nabla \times v_{g}(r,\tau))^{2}\delta(r-r')\delta(\tau-\tau ') \nonumber \\ &+&\frac{mL^{2}}{2}(i\partial_{\tau}v(r,\tau)-\nabla \phi(r,\tau) )^{2} \delta(r -r')\delta(\tau -\tau ')\bigg)
\label{eqn:greensaction}
\end{eqnarray}\end{widetext} 

where $v_{g}(r,\tau):= v(r,\tau)+{1\over m}\nabla \theta(r,\tau)$ is the 
gauge-invariant velocity. The operator $G^{-1}$ (shown in Eqs.~(\ref{eqn:fullgreens1}) - (\ref{eqn:fullgreens2})) is the object of principal computational interest in subsequent sections. Note that besides the field strength contributions and constant offset proportional to $\Delta^{2} $, the complete action can be written as a bilinear form. Using the gauged GP equation~\cite{hohenberg}
and intepreting $\rho(r,\tau)$ as a local condensate density field, it can be shown that the mean-field hydrodynamic effect of the external driving velocity is a depletion of condensate current \cite{balakrishnan,nepo,chelaflores2}: \begin{equation} \partial_{\tau}\rho(r,\tau)-\nabla\cdot(\frac{1}{m}\rho(r,\tau)\nabla\theta(r,\tau))=\nabla\cdot(\rho(r,\tau)v(r,\tau))\end{equation}

In this article, we take the point of view that $\phi(r,t)$ and $v(r,t)$ comprise the gauge-field in the fluid resulting from externally applied driving fields; in particular, the gauge-field is not internally generated by fluctuations. In the nanoaperture array experiment of Hoskinson et al.~\cite{packard}, the oscillations of the hydrodynamic resonator couple to both the condensate atoms and the depletion, like a piston. Thus, the gauge field can be viewed as the externally applied, nonconservative part of the total velocity of the fluid. The response of the phase field to the gauge field is apparent in the Euler and Josephson-Anderson equations that we derive below. An evolving velocity field induces a local chemical potential texture (via the Euler equation) which in turn induces an evolving phase field (via the Josephson-Anderson equation).

\section{\label{sec:effective}Perturbation theory and effective action}

The operator $G^{-1}(r,\tau ;r',\tau ')$ defining the bilinear form in the action Eq.~(\ref{eqn:greensaction}) is \begin{equation}\label{eqn:fullgreens1} G^{-1} = G_{0}^{-1}+G^{-1}_{\dot{\theta}}+G_{v_{g}}^{-1}+G^{-1}_{T} \equiv G_{0}^{-1} +\delta G^{-1} \end{equation} where the individual components of $\delta G^{-1}$ are:  
\begin{eqnarray}\label{eqn:fullgreens2}
G_{0}^{-1} & = & ( \partial_{\tau}-\frac{1}{2m}\nabla^{2}-\mu +V_{0}\Delta ) \delta(r' - r) \delta(\tau - \tau ') \nonumber \\ 
G^{-1}_{\dot{\theta}} &=& [i\partial_{\tau}\theta(r,\tau)+m\phi(r,\tau)] \, \delta(r' - r) \delta(\tau - \tau ') \nonumber\\ 
G_{v_{g}}^{-1} &=& (\frac{1}{2}mv_{g}(r,\tau)^{2})\delta(r' - r) \delta(\tau - \tau ') \nonumber \\ 
G^{-1}_{T} &=& T_{rr'}e^{i(\theta(r',\tau)-\theta(r,\tau))}e^{im\int_{r}^{r'}dr \cdot v(r,\tau)} \delta(\tau - \tau ') 
\label{eqn:G_inverse}
\end{eqnarray} 

Integration over the density field results in a term ${1\over 2} \trlog{G^{-1}}$ in the action. The trace is an integral over all internal positions or momenta and imaginary time $\tau \in [0,\beta]$ arguments. Details of the perturbation expansion for contributions to the action from each part of $\delta G^{-1}$ are given in Appendix A and the general techniques can be found in References~\cite{altland,ambegaokar,korsbakken}.

\subsection{\label{subsection:Delta}Self-consistent equation for $\Delta$}
As mentioned in Section \ref{sec:model}, since there are no space or time derivatives of the Hubbard-Stratonovich fields $\Delta(r,\tau)$, $\Delta^{*}(r,\tau)$ in the action, we can take them to be constant. From the GP equation, $\vert \Delta \vert $ is equal to the density of condensed bosons and we take it to have the same value on both sides of the junction for simplicity. To compute the mean-field value of $\Delta$ from Eq.~(\ref{eqn:greensaction}), we require that it extremizes the action: $\frac{\partial S}{\partial\Delta}=0$. This mean-field is only present in $G_{0}^{-1}$ and in an additional term of quadratic order, and we use the Matsubara frequency and momentum representation of the free inverse Green's function to find the extremum \cite{negele}. Evaluating the resulting Matsubara sum \cite{altland} yields a self-consistent equation for the mean-field $\Delta$ that is analogous to the BCS gap equation:
\begin{equation} \Delta ={1\over V}\int \frac{d^{3}k}{(2\pi)^{3}}\frac{1}{e^{\beta(\frac{k^{2}}{2m}-\mu+V_{0}\Delta)}-1},\end{equation} where $V$ is the volume of the system. In deriving this equation we have assumed that the effect of the gauge field and tunnelling across the aperture contribute negligibly to the mean field value of $\Delta$.

\subsection{\label{subsection:Gtheta}$G_{\dot{\theta}}^{-1}$, Josephson-Anderson equation}

In superconductors, the dynamical (a.c.) Josephson effect is expressed by the Josephson-Anderson equation 
for phase evolution and is dependent on a voltage across the tunnel junction~\cite{devoret_lesHouches}. Although the weakly-interacting Bose gas is not charged, that does not preclude introduction of a 0-component of the gauge field, the fluctuating local chemical potential $\phi(r,\tau)$, that appears in the action with the imaginary-time minimal coupling. The first order contribution of  $\delta G^{-1}\equiv G_{\dot{\theta}}^{-1}$ in Eq.~(\ref{eqn:expansion}) vanishes due to the periodic boundary conditions of $\theta(r,\tau)$ on $[0,\beta ]$ and the requirement that the integral over both sides of the aperture (i.e. both reservoirs) is zero. However, a stationary phase equation with respect to $\phi(r,\tau)$ can be derived from the action by finding the extremum of the ``electric field" energy density term in Eq.~(\ref{eqn:greensaction}), resulting in: \begin{equation}i\del_{\tau}v(r,\tau)=\nabla \phi(r,\tau) \label{eqn:euler}\end{equation}
When transformed into real time, this becomes a classical Euler equation relating the acceleration of the driving velocity to a chemical potential difference across the aperture \cite{ueda}.

The second order term in Eq.~(\ref{eqn:expansion}) gives a nonvanishing contribution: \begin{equation} -\frac{1}{4}\int_{0}^{\beta} d\tau \tilde{G}_{0}(0,\tau ; 0,\tau)^{2}\bigg(i\del_{\tau}\tilde{\theta}(0,\tau)+m\tilde{\phi}(0,\tau) \bigg)^{2} \label{eqn:capacitive} \end{equation} where the tilde signifies a move to momentum space. $\tilde{G}_{0}(0,\tau ; 0,\tau) =: n$ is the ($\tau$-independent) number of $k=0$ bosons so it can be pulled out of the integral, resulting in the coefficient $-{n^{2}\over 4}$ (taking into account also the factor of ${1\over 2}$ multiplying the perturbation series). The imaginary-time Josephson-Anderson equation can then be derived at this order from $\functderiv{S}{\tilde{\phi}(r,\tau)}=0$ together with the global phase and chemical potential configurations for each reservoir, $\theta(r,\tau)=\theta_{R/L}(\tau)$, $\phi(r,\tau)=\phi_{R/L}(\tau)$, yielding: \begin{eqnarray} i\del_{\tau}\theta_{\mathrm{L}}(\tau)&=&-m\phi_{\mathrm{L}}(\tau) \nonumber \\ i\del_{\tau}\theta_{\mathrm{R}}(\tau)&=&-m\phi_{\mathrm{R}}(\tau) \end{eqnarray} 
Subtracting these equations gives the usual form of the 
Josephson-Anderson equation for evolution of the phase difference $\Delta \theta(\tau)$ across a junction:
\begin{equation} i\del_{\tau}\Delta \theta(\tau) = -m \Delta \phi(\tau) \label{eqn:ja}\end{equation} 
We can now use the mean-field Euler equation, Eq.~(\ref{eqn:euler}), derived from the bare theory to write the Josephson-Anderson equation in gauge-invariant form. Noting that $\phi_{L}(\tau)=\phi_{R}(\tau)+i\int_{r_{R}}^{r_{L}}dr \cdot \del_{\tau}v(r,\tau)$, we define the gauge invariant phase-difference by \begin{equation}\label{eqn:gaugeinvphase} \Delta\gamma(\tau) \equiv \theta_{L}(\tau)-\theta_{R}(\tau)+m\int_{r_{R}}^{r_{L}}dr \cdot v(r,\tau) \end{equation} and an analog of the electric field by \begin{equation} \zeta =\int_{r_{L}}^{r_{R}}dr\cdot (-\nabla\phi(r,\tau) +i\del_{\tau}v(r,\tau)) \nonumber \end{equation} A rearrangement of Eq.(\ref{eqn:ja})-(\ref{eqn:gaugeinvphase})
then yields the desired gauge invariant form of the Josephson-Anderson equation as 
\begin{equation} i\del_{\tau}\Delta \gamma(\tau) = -m \zeta(\tau). \end{equation}

To assess the contribution to the effective action $S_{\mathrm{eff}}$, Eq.~(\ref{eqn:capacitive}) should be expressed in terms of $\Delta\gamma(\tau)$. $\phi_{L}(\tau)$ can be eliminated from the action using the Euler equation and $\phi_{R}(\tau)$ can be eliminated by a Gaussian integration (or vice versa, see \ref{subsection:thetaappen}). The result is a capacitive term in the effective action: \begin{equation}S_{C} = \int_{0}^{\beta}d\tau \, E_{C}\bigg(\del_{\tau}\Delta \gamma(\tau) \bigg)^{2} \label{eqn:capterm}\end{equation} where the microscopic expression for $E_{C}$ (from the Gaussian integral) is ${n^{2}V^{2} \over 8}$. Since in our analysis the Euler equation is considered a hard constraint, the electric energy density vanishes and $S_{C}$ is the only contribution from $G_{\dot{\theta}}^{-1}$. Here and in other parts of the single aperture calculations, we neglect cross terms of the form $\trlog{[G_{0} \delta G_{1}^{-1}G_{0}\delta G_{2}^{-1}\cdots ]}$ which are, however, necessary for generating interactions between apertures in the multiaperture case.

\subsection{\label{subsubsection:velocity}$G_{v_{g}}^{-1}$, circulation quantization, Hess-Fairbank equation}

Using Eq.~(\ref{eqn:expansion}) to expand the contribution of $G_{v_{g}}^{-1}$ from Eq.~(\ref{eqn:G_inverse}) to first order results in a term quadratic in the
gauge invariant velocity field $v_{g}(r,\tau)$ (i.e., a massive term for $v_{g}$). In superconductors, the physical consequence of a massive vector field is the Meissner effect, a repulsion of magnetic fields from the interior of the superconductor up to a certain penetration depth which is dependent on the superfluid density~\cite{tinkham}. The analogous effect for $^{4}$He is the Hess-Fairbank effect, in which the superfluid mass density does not respond to rotation of the container due to an energy barrier to vorticity entering the superfluid~\cite{hess}. The massive term for $v_{g}$ in the action suppresses fluctuations of the magnitude of the macroscopic phase gradient ${1\over m}\nabla \theta$ from that of the driving velocity $v(r,\tau)$ in the bulk of the system.  A stationary phase analysis of the action with respect to $v_{g}(r,\tau)$ at this order (see Appendix \ref{sec:vappen}) yields:
\begin{equation} {n\over 2} v_{g}(r,\tau)-L^{2}\nabla^{2}v_{g}(r,\tau)=0, \label{eq:fairbank} \end{equation} 
which is a London equation describing the decay of the gauge-invariant velocity $v_{g}(r,\tau)$ in the interior of the bosonic system, with penetration depth $\lambda = \sqrt{2L^{2}\over n}$.

\begin{figure}
\includegraphics[scale=.6]{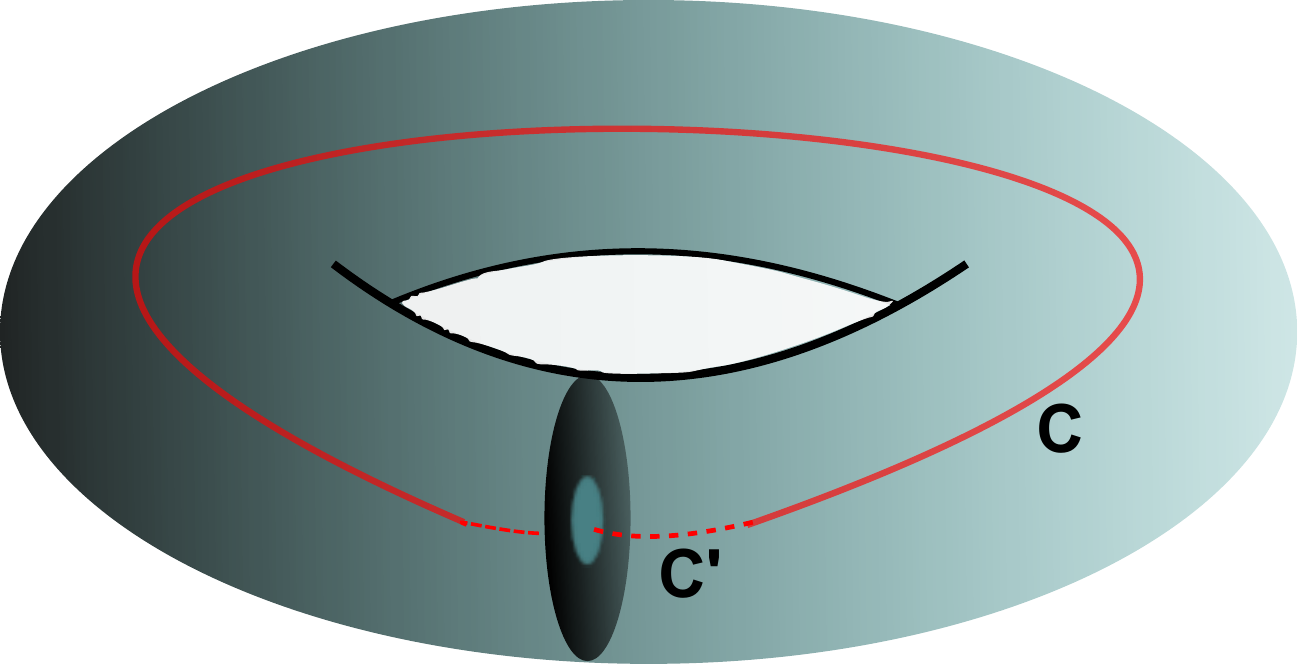}%
\caption{\label{circ}The contours used to derive superfluid fluxoid quantization. $C'$ is threaded through the aperture in the septum.}
\end{figure}

Because we have included an external velocity field, it is useful to explore the consequences of this on circulation quantization. In the absence of driving ($v(r,\tau)=0$), the circulation integral is quantized in values of the circulation flux $\Phi_{0} = {2\pi \over m}$, due to the single-valuedness of the phase: $\oint dr\cdot v_{g}(r,\tau) = {1\over m}\oint \nabla \theta(r,\tau) = {2\pi \ell \over m}$, $\ell \in \mathbb{Z}$. It seems clear that some form of the quantization should carry over to the driven case. To this end, we will integrate the London equation over a properly chosen contour. By 
analogy with Amp\`{e}re's law, we interpret $ \curl{\omega(r,\tau)} $ as a current $j(r,\tau)$. We take a line integral of the London equation (\ref{eq:fairbank}) around the torus on a path $C$ which goes all the way around the torus, except for a missing segment $C'$ across the aperture (Figure \ref{circ}) \begin{equation} \int_{C}dr\cdot \bigg(j(r,\tau)+\frac{n}{2L^{2}}(v(r,\tau)+\frac{1}{m}\nabla\theta(r,\tau))\bigg) =0 \nonumber \end{equation}

Since $\theta$ is only defined mod $2\pi$, one must have $\displaystyle \int_{C}dr\cdot \nabla\theta(r,\tau)=2\pi l -\Delta\theta(\tau)$, $l\in\mathbb{Z}$ where $\Delta\theta$ is the line integral of the phase over $C'$ (local phase difference across the aperture). The line integral of the driving velocity gives the external circulation $\Phi$ 
\begin{equation} \int_{C}dr \cdot v(r,\tau)+\int_{\mathrm{C'}}v(r,\tau)\cdot dr=\int_{A}\omega \cdot dS = \Phi \end{equation}
Combining these two equations gives the generalized circulation quantization condition: 
\begin{equation} 
\label{eq:solve} 
\frac{1}{\Phi_{0}}\bigg(\Phi+\frac{2L^{2}}{n}\int_{C}dr\cdot j(r,\tau) \bigg) = \frac{\Delta\gamma (\tau)}{2\pi}-\ell 
\end{equation} 
with $\ell \in \mathbb{Z}$ and $\Delta\gamma(\tau)$ the gauge-invariant phase difference in Eq.(\ref{eqn:gaugeinvphase}). $\Phi_{0}$ is the circulation quantum, $\Phi_{0}={2\pi \over m}$. At distances into the bulk superfluid greater than the penetration depth, $j(r,\tau)=0$ so that the above equation reduces to an equation for the quantization of the circulation due to the driving velocity. 
The general form of Eq.~(\ref{eq:solve}) expresses quantization of the superfluid ``fluxoid" \cite{degennes} which contains contributions from vorticity due to the driving current in addition to the superfluid circulation. 

It remains to determine the contribution of $v_{g}$ to the effective action for the gauge-invariant phase difference. Inclusion of the vorticity (circulation energy density) in the bare action Eq.~(\ref{eqn:greensaction}) and requiring that the London equation hold results in the cancellation of the first order contribution of $G_{v_{g}}^{-1}$ by the circulation energy density (see Appendix \ref{sec:vappen} for derivation). It should be mentioned that in deriving this cancellation, we ignore a topological surface term $\int_{\partial V} v_{g} \wedge \omega$. In fact, had we included in the Higgs action source terms for $v_{g}(r,\tau)$ and $\omega(r,\tau)$, parametrized the vortex current by an appropriate gauge field, and integrated out $v_{g}(r,\tau)$, the effective theory for the phase texture and vortex gauge field would be a $BF$ topological field theory \cite{hansson,thouless}. In this work, we do not consider explicitly the dynamics of vortices (but see discussion in Section~\ref{sec:conclusion}).

The second order contribution of $G_{v_{g}}^{-1}$ gives a nonvanishing contribution to the effective action for $\Delta \gamma$ and simplifies to: \begin{widetext} \begin{equation} -{m^{2}\over 16}\int d^{3}kd^{3}k'  \, n_{k}n_{k'}\int {d^{3}qd^{3}\xi \over (2\pi)^{6}}\,  \tilde{v}_{g}(q,\tau)\tilde{v}_{g}(k-k'-q,\tau)\tilde{v}_{g}(\xi ,\tau)\tilde{v}_{g}(k'-k-\xi,\tau) \end{equation} \end{widetext} This expression is a convolution in momentum variables, but can be approximated as local in momentum because $n_{k}$ is exponentially suppressed for $k\neq 0$ at low temperatures. Because the quadratic term in $v_{g}$ gives rise to a linear term in $\Delta\gamma$ (Eq. (\ref{eqn:A5})) the term quartic in $v_{g}$ results in a quadratic term for $\Delta\gamma(\tau)$: \begin{equation} S_{Q}= -E_{Q}(\ell - {\Delta\gamma(\tau) \over 2\pi})^{2} \label{eqn:quad} \end{equation} where $E_{Q} = m^{2}n^{2}\Vert v \Vert^{2}L^{4} \Phi_{0}^{2}/16$. 

We note here that in the present analysis of driven bosonic flow through an aperture, $G_{v_{g}}^{-1} $ is strictly second order in the gauge-invariant velocity $v_{g}$, while the corresponding perturbative contribution for superconducting current flow through a Josephson junction also contains a term linear in $v_{g}$~\cite{ambegaokar,korsbakken}. The consequence is that our second order expansion in $G_{v_{g}}^{-1}$ is quartic in $v_{g}$. This difference is a result of the polar decomposition of the bosonic fields made here into real components, in contrast to the superconducting case in which one must work with Nambu spinors. Expanding the square in Eq.~(\ref{eqn:quad}) shows that the effective action has both quadratic and linear dependence on $\Delta \gamma$. The latter will result in a quantized constant term (a quantized current-bias) in the current-phase equation while the former will give a term proportional to $\Delta\gamma$ (see Section \ref{sec:josephson}).

\subsection{$G_{T}^{-1}$, periodic potential}\label{sec:tunneling}
We require that the tunneling matrix $T_{rr'}=0$ when $r$ and $r'$ are on the same side of the aperture and, for simplicity, $T_{rr'}=T=\mathrm{const.}$ when $r$ and $r'$ are on opposite sides of the aperture. In the perturbation expansion, we must integrate over all possible positions which give nonzero tunneling matrix elements. The resulting term in the effective action is: \begin{equation}  S_{J} = Tn\int_{0}^{\beta }d\tau \cos{\Delta\gamma(\tau)}\nonumber \label{eqn:tunnel}\end{equation} where $n$ is the zero-momentum occupation. If the perturbation expansion is continued and the imaginary-time integrations are approximated by a single one, higher harmonics of the $\cos(\Delta\gamma(\tau))$ interaction result; we will not include these in our analysis. These interactions can be shown to be of less relevance than the leading interaction (decreasing faster as the high-energy cutoff is lowered) by background field RG methods \cite{wen}. However, if one keeps imaginary-time arguments distinct (i.e., preserves time non-locality) in the second order contribution, the second order term may be included as a dissipative contribution to the effective action (see Section \ref{sec:tunnappen}). AES use an analogous term of this order to model the effect of quasiparticle-macroscopic phase difference scattering on the current in the Josephson junction.

\section{\label{sec:josephson}Effective action and current-phase relations}
\subsection{\label{sec:effectiveaction}Effective action}
\begin{figure}
\includegraphics[scale=.5]{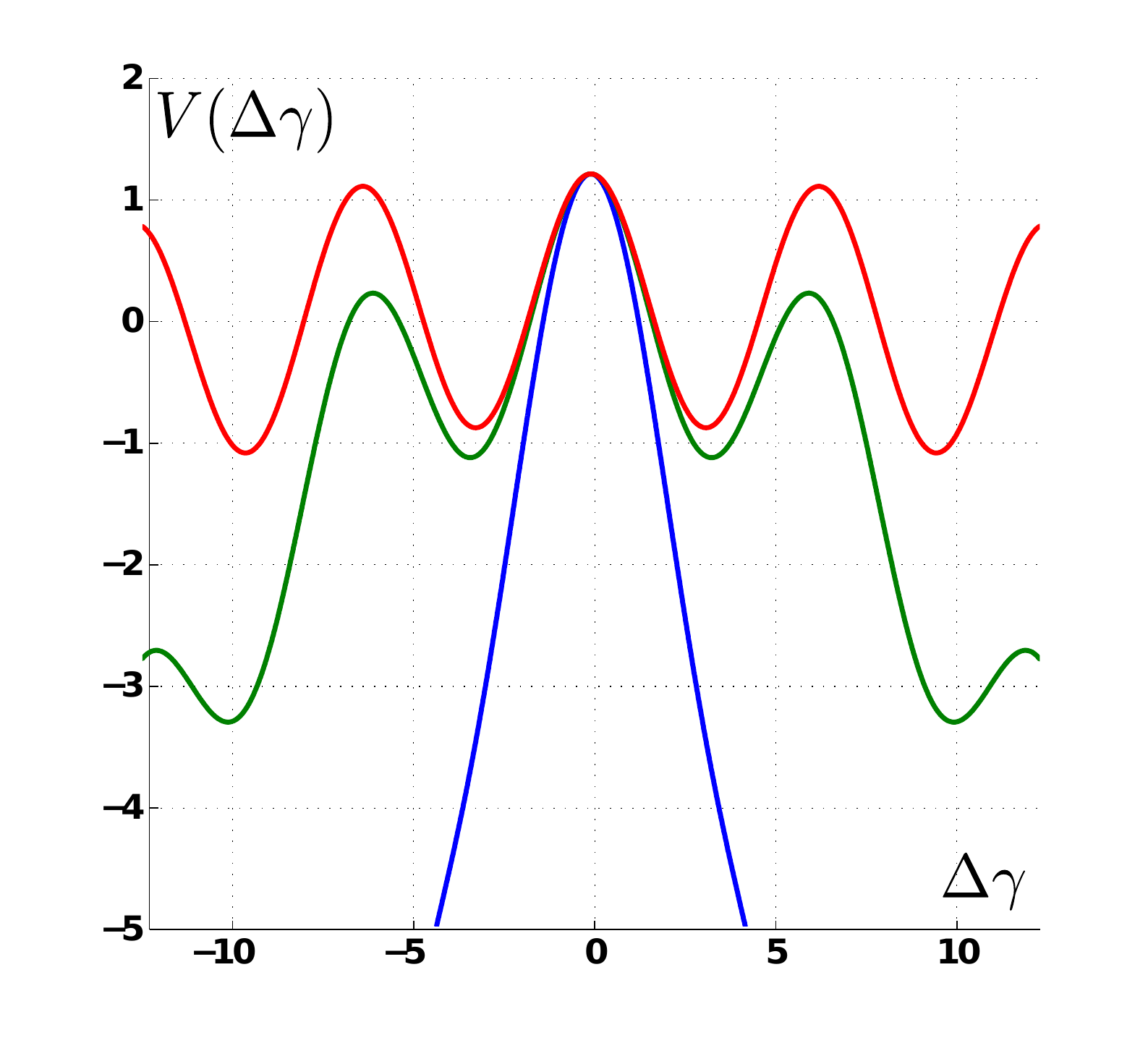}%
\caption{\label{pot}
 The potential of the effective Euclidean action, $V[\Delta\gamma] = - E_{Q}(\ell - {\Delta\gamma/2\pi})^{2} + E_{J} \cos{\Delta\gamma}$ versus $\Delta \gamma$ shown with $\ell = 0$ for $E_{J}=1$ and three values of the ratio ${E_{Q}/E_{J}}$: ${E_{Q}/E_{J}}$ = 10 (blue), 1 (green), 0.1 (red). Evaluating the potential at $\ell \neq 0$ breaks the $\Delta\gamma \rightarrow -\Delta\gamma$ symmetry.  }
\end{figure}

The effective action for the gauge-invariant phase difference is determined from Eq.~(\ref{eqn:capterm}), Eq.~(\ref{eqn:quad}), Eq.~(\ref{eqn:tunnel}) to be $S_{\mathrm{eff}}=S_{C}+S_{Q}+S_{J}$. Explicitly: \begin{widetext}\begin{eqnarray} S_{\mathrm{eff}}[\Delta\gamma(\tau); l,\beta]&=& \int_{0}^{\beta}d\tau \,  E_{C}\bigg(\del_{\tau}\Delta \gamma(\tau) \bigg)^{2} - E_{Q}(\ell - {\Delta\gamma(\tau) \over 2\pi})^{2} +E_{J} \cos{\Delta\gamma(\tau)} \label{eqn:effective}\end{eqnarray}\end{widetext}
 where the microscopic expressions for the coefficients have been derived above: $E_{C}= {n^{2}V^{2}/8}\; , E_{Q} = {m^{2}n^{2}\Vert v \Vert^{2}L^{4}/16}\; , E_{J}= Tn$. 
This effective action describes a particle on a ring with a potential that is a sum of a parabolic and cosine terms, i.e., 
\begin{equation}\label{eqn:V_Dgamma}
V[\Delta\gamma] = -E_{Q}(\ell - {\Delta\gamma(\tau)/2\pi})^{2} + E_{J} \cos{\Delta\gamma(\tau)}.
\end{equation} (see Figure \ref{pot}). In the partition function involving $S_{\mathrm{eff}}$, the sum over $\ell \in \mathbb{Z}$ counts the winding number of the macroscopic phase. Changes in $\ell$ correspond to phase slips across the aperture. The behavior of the effective potential for $\ell = 0$ and a range of relative values of the parameters $E_{J}$,$ E_{Q}$ is shown in Figure \ref{pot}. Note that for a given $\ell$ there is an infinite number of local minima of the potential. The generalized circulation quantum condition, Eq.~(\ref{eq:solve}), may be used to further write the effective action solely in terms of the circulation $\Phi$. The Hamiltonian corresponding to this action is formally similar to that used to describe \textit{rf} SQUIDs and superconducting flux qubits\cite{korsbakken,wendin,devoretwalraffmartinis} and has been used previously to analyze coherent quantum phase slips \cite{manucharyan}.

The quadratic contribution of $\Delta\gamma$ in the effective potential differentiates this potential from the sinusoidal-plus-linear or ``washboard" form of effective potential found for a current-biased Josephson junction \cite{wendin}. The effective action derived here for driven bosonic flow through an aperture differs from that derived by AES for superconducting flow through a Josephson junction in two respects.  First, for the driven bosonic flow, the gauge field contribution $G_{v_{g}}^{-1}$ to the effective action at second order in perturbation theory is quartic in $v_{g}$, resulting in a term quadratic in $\Delta\gamma$ and hence a parabolic contribution to the potential. In contrast, the contribution from the superconducting superfluid velocity to the effective action for a superconducting tunnel junction is linear in the phase difference variable (see Eq. (31) in Ref.~\cite{ambegaokar}) and second order terms arise only from the additional inductive energy.   Second, we have neglected the second order tunneling perturbation which is nonlocal in time: inclusion of this would, by analogy with the analysis of AES, give rise to dissipation in the aperture array.

In Section \ref{sec:rg} below we will analyze the temperature-dependence of $E_{J}$.  Because the temperature-dependence will enter through the ratio of $E_{Q}$ to $E_{C}$ in Eq.~(\ref{eqn:effective}), we now show that the latter ratio can be written in terms of the ratio of two characteristic lengths of the system. According to the analysis above: \begin{eqnarray} {E_{Q}\over E_{C}} &=& {2\pi^{2}\Vert v\Vert^{2}L^{4} \over V^{2}} \nonumber \\ &=& {2\pi^{2}\Vert v\Vert^{2}n^{2}\lambda_{L}^{2} \over 2V^{2}},  \end{eqnarray} 
where $\lambda_{L}$ is the penetration depth of the gauge invariant velocity $v_g$ (Eq.~\ref{eq:fairbank}).
The condensate density, ${n/ V}$, is related to the healing length at nonzero $T$ in the Popov theory by ${n/V}={1/8\pi a \xi(T)^{2}}$, where $a$ is the s-wave scattering length \cite{pethick}. We can then express the ratio by 
\begin{equation}\label{eqn:scale} 
{E_{Q}\over E_{C}}= {\Vert v\Vert^{2}\over (2a \sqrt{2})^{2}} \bigg({\lambda_{L} \over 2\xi(T)} \bigg)^{4}
\end{equation} 
We will use Eq.(\ref{eqn:scale}) to analyze the current-phase relation of a single aperture in Section \ref{sec:singap} below.

\subsection{\label{subsection:gencurrphase}Generalized current-phase relations}
The current-phase relation is the central equation of weak link Josephson phenomena.  Not only does it reflect the macroscopic quantum nature of the flow through an aperture, but it can also yield indirect information on the microscopic dynamics of the constituent particles at the aperture.  The most extensive studies of such flow for interacting bosons in atomic systems (as opposed to superconductors) have been made for $^4$He, where experiments with driven flow through nanoaperture arrays reveal the existence of two different current-phase relations in different temperature regimes below the $\lambda$ point.  We will use the microscopically derived effective action for the single aperture, Eq.~(\ref{eqn:effective}) to construct the current-phase relation for the weakly interacting bosonic system and use it to analyze the $^{4}$He flow experiment, bearing in mind that liquid $^{4}$He is a strongly interacting system so the analysis remains qualitative.  Since observation of Josephson effects under external driving has not yet been observed for weakly interacting Bose condensed gases, although both Josephson coupling~\cite{oberthaler} and persistent flow~\cite{ryu} have been observed in different geometries (double well and toroidal traps, respectively), we also expect that our analysis will be applicable to driven condensed weakly-interacting Bose gases separated by aperture arrays.

\subsubsection{\label{sec:singap}Current-phase equation for single aperture}

The current-phase equation resulting from the effective action $S_{\mathrm{eff}}$ is obtained as the stationary phase equation $\delta S_{\mathrm{eff}} / \delta \Delta\gamma (\tau)=0$.  This equation is derived in convenient form by first defining the ``density difference" field $\Delta n(\tau)$ that is canonically conjugate to $\Delta \gamma (\tau)$, by Legendre transformation of the Lagrangian in the path-integral. Specifically, the kinetic term of Eq.~(\ref{eqn:effective}) is changed via:

\begin{widetext}\begin{equation}
e^{-\int_{0}^{\beta}d\tau \, E_{C}\bigg(\del_{\tau}\Delta\gamma(\tau) \bigg)^{2} }\propto \int \mathcal{D}[\Delta n(\tau)]e^{-\int_{0}^{\beta}d\tau \, {1 \over 4E_{C}}\Delta n(\tau)^{2} + i\int_{0}^{\beta}d\tau \,\Delta n(\tau)\del_{\tau}\Delta\gamma(\tau)} 
 \label{eqn:Legendre}
\end{equation} \end{widetext}

Performing a stationary phase analysis with respect to $\Delta\gamma (\tau)$ on the resulting Legendre transformed Eq.~(\ref{eqn:effective}) then yields the general imaginary time current-phase equation
\begin{eqnarray}i\del_{\tau}\Delta n(\tau)-E_{J}\sin (\Delta\gamma(\tau))  +{E_{Q}\over 2\pi^{2}}\Delta\gamma(\tau)+{ E_{Q}\ell \over \pi} =0 \label{eqn:josephson}
\end{eqnarray} 
The current-phase relation in terms of the real-time current $I(t)= {d\Delta n / dt}$ is obtained by a Wick rotation of Eq.~(\ref{eqn:josephson}). The term linear in  $\Delta\gamma$ confirms that this current-phase relation constitutes an analog for weakly interacting condensed bosons of the generalized Josephson equation for an \textit{rf} SQUID. The current-bias part of the current-phase relation is constant and quantized, proportional to $\ell \in \mathbb{Z}$. We emphasize that Eq.~(\ref{eqn:josephson}) contains all terms necessary to describe a linear-to-sinusoidal current-phase transition.

The different forms of the current-phase relation in different physical regimes correspond to specific values of the parameters $E_{Q}$ and $E_{J}$. For $E_{J}=0$, the current-phase relationship of Eq.~(\ref{eqn:josephson}) is linear and corresponds to the small amplitude oscillations of a pendulum~\cite{raghavan,leggett}. However, the effect of an $\ell$-dependent current-bias persists. For $E_{Q}=0$, this equation reduces to the imaginary-time version of the Josephson equation, with critical number current equal to $E_{J} = Tn$ (and mass current given by $m E_{J}$).

To determine (in imaginary time) the classical equation for $\Delta \gamma$, one can require the exponent of Eq.~(\ref{eqn:Legendre}) to be stationary with respect to variations in $\Delta  n$. This results in the relation $i\del_{\tau}\Delta\gamma = {-1\over 2E_{C}}\Delta n $, analogous to $m\dot{q} = p$ in classical mechanics. Substituting this relation into Eq.(\ref{eqn:josephson}), it is then evident that
for $E_{Q}\ll E_{C}$ the quantized current-bias and the coefficient of the linear term are negligible; it is in this regime that purely sinusoidal oscillations should be observed.
In this limit, one recovers the imaginary time version of the classical (fixed length) pendulum equation with amplitude ${E_{J}/ 2E_{C}}$, i.e.,
\begin{eqnarray}
\del^2_{\tau} \Delta \gamma = \frac{E_{J}}{2E_{C}  } \sin (\Delta \gamma ),
\label{eqn:pendulum}
\end{eqnarray}
which constitutes a well-known classical analogue of the Josephson effect~\cite{leggett}.

\begin{figure}
\begin{center}
\includegraphics[scale=.7]{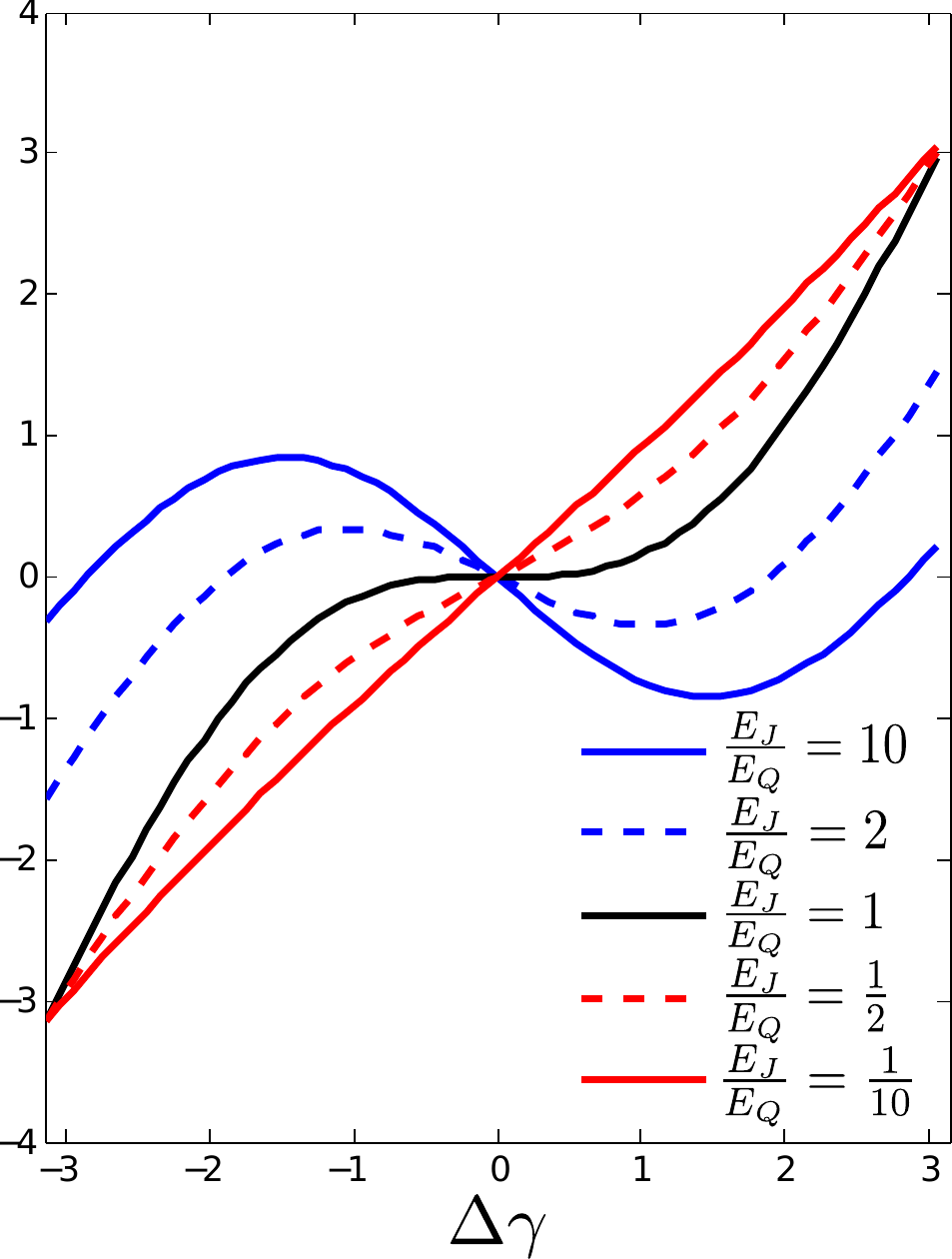}%
\caption{\label{current_phase} Plots of real time ($i{d\over d\tau} \rightarrow -{d\over dt}$ in Eq.~(\ref{eqn:josephson}) ) current-phase relations for mixed linear-sinusoidal contributions over $[-\pi ,\pi]$ for $\ell = 0$ and a range of values for the parameter ratio $E_{J}/E_{Q}$ at fixed temperature.  These are to be compared with the experimentally observed current-phase relations shown in Figure 3 of Reference~\cite{packard}. In Section \ref{sec:rg} we derive the temperature-dependence of the current-phase relationship.}  
\end{center}
\end{figure}

More generally, the current-phase relation, Eq.~(\ref{eqn:josephson}), interpolates between two regimes of purely linear and sinusoidal current-phase equations at $E_{J}=0$ and $E_{Q}=0$, respectively (plotted in real time in Figure \ref{current_phase}). These two limiting current-phase behaviors were observed for different temperature intervals in the $^{4}$He nanoaperture array experiments of Ref.~\cite{packard}.

If we set $\lambda_{L}=\lambda_{\mathrm{ap}}$ in Eq.~(\ref{eqn:scale}), with $\lambda_{ap}$ the diameter of a single aperture (see Section \ref{sec:rg} for justification), we may relate the ratio $E_{Q}/E_{C}$ to the ratio $\lambda_{ap}/ 2\xi(T)$ of the characteristic aperture size to the temperature-dependent healing length. As the temperature is decreased, the healing length becomes smaller than the aperture size and the ratio ${E_{Q}/E_{C}}$ grows quartically. Thus if $E_{J}$ is considered fixed, the linear term in the current-phase equation Eq.~(\ref{eqn:josephson}) becomes dominant for low $T$. In contrast, at higher temperatures, e.g. large enough that the healing length is larger than the aperture size, the sinusoidal term would become dominant. Whether thermal fluctuations of the gauge-invariant phase difference wash out the sinusoidal part of the current-phase relation as $T_{\lambda}$ is approached from below depends on the size of $E_{J}$, the scaling of which is derived in terms of $E_{C}$ and $E_{Q}$ in Section \ref{sec:rg}.
 
This qualitative analysis shows that as the temperature is increased towards $T_\lambda$, there can be a transition from a linear current-phase relation at low temperatures to a sinusoidal current-phase relation at higher temperatures (but still below $T_\lambda$).

\subsubsection{\label{sec:twoap}Current-phase relation for two-aperture array}
Within the framework of this theoretical analysis, adding an additional aperture is straightforward and results in a substantially richer set of current-phase phenomena. We analyze here just the two-aperture case, leaving the extension to arrays with large numbers of apertures for future investigation.  We may assume the cross-sectional areas of the two apertures are identical. There are now two tunneling matrices $T^{(1)}_{r,r'}$ and $T^{(2)}_{r,r'}$; we require that $T^{(1)}_{r,r'}$ is nonzero only when $r$ and $r'$ are on opposite sides of aperture 1 \textit{and} both are in a small vicinity of the aperture (similarly for $T^{(2)}_{r,r'}$). In addition to the sum of single aperture effective actions for the gauge invariant phase differences $\Delta\gamma^{(1)}(\tau)$ and $\Delta\gamma^{(2)}(\tau)$, which have been derived in Section~\ref{sec:tunneling}, there is now also a tunneling cross-term that appears at second order in the perturbation theory. This tunneling cross-term generates an effective aperture interaction that may be expressed in terms of the microscopic phase differences across the individual apertures. In particular, with the tunneling amplitudes assumed to be the same, this term adds an interaction to the effective action for two apertures of the form
\begin{equation} S_{\mathrm{int}}= -E_{J}^{2}\cos(\Delta\gamma^{(1)}(\tau))\cos(\Delta\gamma^{(2)}(\tau)).\label{eqn:2aperturecoupling} \end{equation} 
For small phase-differences, expansion of this equation implies that the homogeneous part of degree 2 renormalizes the quadratic parts of the uncoupled contributions to the action and introduces a coupling $\Delta\gamma^{(1)}(\tau)\Delta\gamma^{(2)}(\tau)$, while the homogeneous part of degree 4 introduces a coupling $\Delta\gamma^{(1)}(\tau)^{2}\Delta\gamma^{(2)}(\tau)^{2}$ as well as quartic local potentials for the phase differences. Neglecting these higher order terms, the interaction results in coupled modified Josephson equations which describe classical coupled pendula.

We can use the two-aperture coupling term Eq.~(\ref{eqn:2aperturecoupling}) to rationalize the experimentally observed transition from a linear to sinusoidal current-phase relation in a multi-aperture array. Because the coefficient of the interaction just derived is the square of $E_{J}$ we know that if $E_{J}$ is large compared to $E_{Q}$, the current-phase relation for each individual aperture is approximately sinusoidal \textit{and} that the energy cost for having an inter-aperture phase difference of $\pi$ is $2E_{J}^{2}$. This means that for $E_{J}\neq 0$, it is favorable for the difference of the phase-differences to be $0 \text{ mod } 2\pi$. Hence the amplitude of the oscillation coming from the independent terms is doubled. This is consistent with both the experimental observations of phase difference synchronization as the current-phase relation becomes sinusoidal, i.e. Josephson-like, as well as with the observed linear scaling of the Josephson oscillation amplitude with number of apertures~\cite{packard}. 

\section{\label{sec:rg}Renormalization group analysis for small $E_{J}$}
To make contact with experiment and to justify the qualitative argument presented in Section \ref{sec:singap}, it is desirable to understand how the current-phase relationship of the effective theory, Eq.~(\ref{eqn:josephson}), and in particular the critical current $E_J$, depends on temperature. This can be done by employing RG methods in the small $E_{J}$ regime and 
analyzing the corresponding beta function \cite{rgnote}. The sign of this function determines how the coupling constant $E_{J}$ behaves (i.e. decreases or increases) at low energies/long length scales.
 
Full details of the RG calculations are included in Appendix B. Here we summarize only the key features of this calculation and the results that are relevant to understanding the temperature dependence of the current-phase relation presented in Section~\ref{subsection:gencurrphase} above. We note that in order for the system to be described by the phase-difference only, we must implicitly assume a high-energy cutoff $\Lambda$, beyond which energy scale the effective theory is invalid. At the energy scale determined by $b={\Lambda \over \lambda}$, with $\lambda$ a lower energy scale 
(i.e., $b\in [1,\infty )$), $E_{J}(b)$ is the critical current of the current-phase relation and its magnitude relative to $E_{Q}$ will determine the Josephson character of the current-phase relation.

\begin{figure}
\begin{center}
\includegraphics[scale=.8]{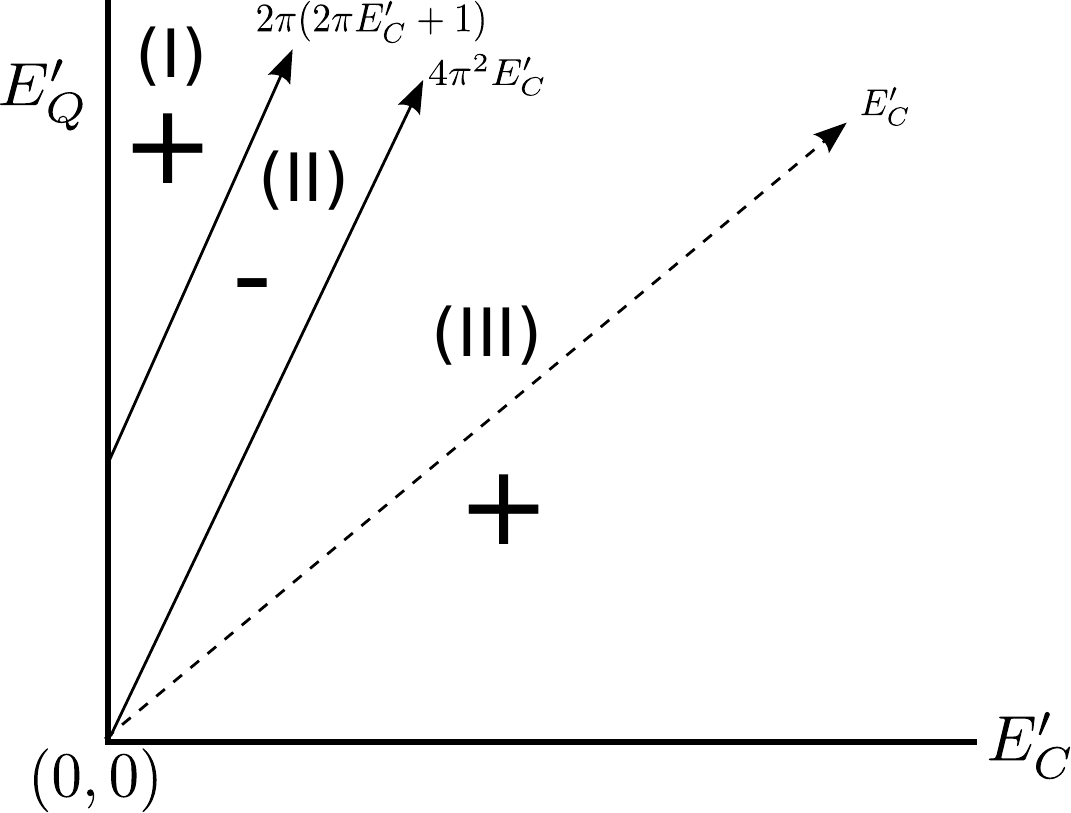}%
\caption{\label{rg} Renormalization group diagram for the Josephson (sinusoidal) contribution to the generalized current-phase relation, Eq.~(\ref{eqn:josephson}).
The solid line separating regions I and II is the marginal line $E_{Q}'=2\pi \left(2\pi E_{C}' +1\right)$ and the solid line separating regions II and III is the singular line ${E_{Q}'=4\pi^{2}E_{C}'}$.  A plus (minus) sign in a given region indicates the direction of the $E_{J}$ RG flow, corresponding to the cosine potential being relevant and increasing with increasing $b$ (irrelevant and so decreasing with increasing $b$). A relevant periodic potential results in a Josephson term in the current-phase relation.  Each temperature $T$ defines a ray in this positive quadrant (see text).   }
\end{center}
\end{figure}

Since we are concerned here with the scaling of $E_{J}(b)$, we neglect the scaling of $E_{C}$ and $E_{Q}$. If $E_{J}(b)$ decreases (increases) as we consider low energy scales, we infer that the low-energy current phase relation Eq.~(\ref{eqn:josephson}) does not contain (does contain) a sinusoidal term. The resulting beta function is then given by
\begin{equation}\label{eqn:beta} 
\beta(E_{J})\equiv b{dE_{J}\over d b}\vert_{ b=1}=\bigg( 1+{2 \pi \Lambda\over 4\pi^{2}E_{C}\Lambda^{2}-E_{Q} }\bigg)E_{J}. \nonumber  
\end{equation} Integrating this differential equation by separating variables and transforming to dimensionless parameters (using the naive scaling dimension of each) $E_{Q}'={E_{Q}\over \Lambda}$, 
$E_{C}'=E_{C}\Lambda$, yields the following scaling field for $E_{J}$: 

\begin{eqnarray} 
E_{J}(b) \propto E_{J}b^ {\left( {1+g } \right)},  \nonumber \\
g = {1 \over {2\pi \left( E_{C}' - {E_{Q}'\over 4\pi^{2}} \right) }}
\end{eqnarray} 
We have confirmed the validity of this scaling field with a background RG calculation~\cite{wen}. When the exponent $1+g$  is negative,  $E_{J}(b)$ will be irrelevant and disappear at low energies, while when the exponent is positive $E_{J}(b)$ is relevant and grows at low energies. 
Figure \ref{rg} shows the resulting RG flow diagram for $E_{J}$ in the positive $\left( E_Q', E_C' \right)$ quadrant.

There are two important features in this RG diagram for $E_{J}$.  
First, the singular line defined by ${E_{Q}'=4\pi^{2}E_{C}'}$ (where the denominator of $g$ goes to zero) and second, the marginal line at $E_{Q}'=2\pi \left(2\pi E_{C}' +1\right)$ (where $1+g =0$).  
We can analyze the singular line in terms of the ratio ${E_{Q}/E_{C}}$ considered at the beginning of Section \ref{sec:josephson} (see Eq.~(\ref{eqn:scale})). In order to evaluate this ratio as a function of the renormalization scaling $b$, we must choose a value for the high-energy cutoff, $\Lambda$. In the 
low temperature helium nanoaperture array experiments of Ref.~\cite{packard}, the largest energy scale 
is the kinetic energy of the driving velocity. We therefore employ an energy cutoff value $\Lambda = {\Vert v \Vert \over 2a\sqrt{2}}$. Returning to Eq.~(\ref{eqn:scale}), we see that the condition for the singularity will then occur at a temperature $T_{1}$ such that \begin{equation}{E_{Q}'\over E_{C}'} = \bigg({ \lambda_{L} \over 2\xi(T_{1}) } \bigg)^{4} =4\pi^{2} \label{eqn:scaled} \end{equation} 

In the following analysis of the experiment in Ref.\cite{packard}, we set $\lambda_{L}=\lambda_{\mathrm{ap}}$, with $\lambda_{\mathrm{ap}}$ the diameter of an aperture in the array. This is consistent with analysis of the first critical angular velocity for appearance of vortex lines in rotating annular reservoirs of liquid $^{4}$He, where the annular width appears in the expression for critical angular velocity in an identical form to the London penetration depth for first critical magnetic field in type-II superconductors  \cite{bendt1,bendt2}.
From the known temperature-dependence of the healing length $\xi(T)$ for He II~\cite{healing}, and using the experimental aperture diameter of 40 nm, this scaling singularity is found to occur at a critical temperature $T_{1}\approx T_{\lambda}-20 \mathrm{mK}$. Conversely, for a general temperature $T$, the right-hand side of Eq.~(\ref{eqn:scaled}) shows that each value of $T$ defines a ray in the coupling-constant space $(E_{C}',E_{Q}')$, and specifying the exact point in coupling constant space requires knowledge of either $E_{C}'$ or $E_{Q}'$. 

We now consider the nature of the current-phase relation for temperatures below and above $T_{1}$. Figure~\ref{rg} summarizes the structure of the low-energy current-phase relation in the $(E'_{Q}, E'_{C})$ plane for the regime of small Josephson coupling $E_{J}$. At temperatures below $T_{1}$, $E_{Q}'/E_{C}' > 4\pi^2$ and we are either in region I or region II of Figure \ref{rg}. In the former case we might expect a mixed sinusoidal/linear flow, while in the latter case we expect only a linear current-phase relation. In region I, $E_{Q}'$ is always nonzero so some linear flow is always present. Although the experiment in Ref.~\cite{packard} does not address this particular temperature regime, we can use a number of arguments to predict the expected balance between linear and sinusoidal contributions as a function of temperature within this regime. At low temperature when the number of condensed bosons is large, or whenever the tunneling amplitude is very large or very small, the $E_{J}\cos \Delta \gamma$ part of the action can be treated using the Villain approximation~\cite{kleinert} which would renormalize $S_{Q}$ and lead to a purely linear current/phase equation. In support of this argument is the fact that for $E_{Q}'>2\pi$ and constant, the value of $E_{C}'$ is lower in region I than in region III. A low value of $E_{C}'$ implies a high energetic cost for density difference fluctuations (see Eq.~(\ref{eqn:Legendre}). Since the density difference is canonically conjugate to the gauge-invariant phase difference, we expect that a low variance in the value of the former quantity allows for a high variance in $\Delta\gamma $ and hence for the Josephson flow contribution to the current-phase relation to be washed out.  

At temperatures above $T_{1}$, the periodic potential is relevant (region III). The current-phase relationship, Eq.~(\ref{eqn:josephson}) will always have a nonvanishing contribution from sinusoidal flow in this regime (while the system remains below $T_{\lambda}$ although it may be mixed with linear flow. For $E_{Q}'$ small, nearly pure Josephson oscillations should be observed.

To support the validity of this analysis of the small $E_J$ current-phase relation, we place two results from the experiments of Ref.~\cite{packard} that exhibit different current-phase behaviors into the context of the RG diagram, Figure~\ref{rg}. For example, at $T_{\lambda}-T = 27$mK a linear current-phase relation is observed. Employing the experimental formula for the healing length~\cite{healing} and an aperture width $\lambda_{ap} = 40$nm, yields the ray ${E_{Q}'/ E_{C}'}\approx 94$ for this temperature. Since the experimental current-phase relation has linear character at this temperature, we expect that this point lies in region II below the $E_{Q}=2\pi(2\pi E_{C} +1)$ line. The second point we analyze is $T_{\lambda}-T = 0.8$ mK.  Here the experiment shows nearly pure Josephson oscillations and experimental estimates for healing length and aperture width yield the ray ${E_{Q}'/ E_{C}'}\approx 8.0 \times 10^{-3}$.  Consequently this higher temperature point lies in region III, far below the $E_{Q}'=E_{C}'$ line and in a region where $E_{Q}'$ is negligible.

We emphasize that pure sinusoidal Josephson oscillations (without the modified dynamics due to parabolic potential) should be found in region III of Fig.(\ref{rg}) only. This is a regime of considerable interest for applications of Josephson phenomena in liquid $^{4}$He to metrology~\cite{packardnew} and for development of 
circulation analogues of superconducting flux qubits~\cite{tian,bpanderson}. The experimental challenge in accessing this regime lies in the fabrication of small enough nanoaperture arrays in order for the $E_{Q}'=4\pi^{2}E_{C}'$ line to be reached deep in the condensed phase and not near the critical point. 

\subsection{\label{sec:rg_multiple} Multiple apertures}

The present analysis is made for a single aperture. Observing a Josephson current for a bosonic superfluid in a single driven nanoaperture is known to be a challenging task, due to the small amplitude of oscillation compared to the amplitude of oscillations of the driving device. Our analysis shows that if the healing length of an interacting Bose gas can be made over twice the characteristic aperture size, nearly pure Josephson oscillations would be observable. Unfortunately, for driven liquid $^{4}$He in aperture arrays of $\lambda_{ap}\sim 40$nm, the system for which all such experiments have been performed to date, this regime is nearly precluded by the lambda transition. In superfluids with larger zero-temperature coherence lengths (e.g., the paired fermion superfluids, including $^{3}$He and many type-II superconductors) the Josephson effect is consequently more robust with a single aperture.

For bosonic superfluids such as liquid $^{4}$He and trapped dilute Bose gases, it is of interest to consider what changes to the present analysis are required by having multiple apertures.  If tunneling amplitudes at each aperture are the same and each aperture has the same size and shape, even the particulars of the weak $E_{J}$ coupling RG calculation should carry over. The most important change in going from one aperture to multiple apertures is the presence of the phase-difference interaction and the independent tunneling terms as mentioned in Section \ref{sec:twoap}. If the phase-difference interaction favors a uniform value, the classical configurations will be phase-locked, independent tunneling terms will add up and the overall tunneling amplitude will be scaled by $M$, with $M$ the number of apertures in the array. Consequently, the amplitude of the Josephson oscillation is multiplied by $M$ and it is easier to observe. It should be noted that the presence of multiple apertures introduces new, higher-order operators in the effective action. In general, their anomalous scaling dimensions (and hence their operator relevance) are different from  that of the $\cos(\Delta\gamma)$ potential.

\section{\label{sec:conclusion}Conclusion}
In this work we have derived and analyzed an effective theory of gauge-invariant phase differences across simple aperture arrays starting from a local U(1) gauge theory. The stationary-phase approximation to the local U(1) gauge theory at first and second order expansion of the one-loop contribution to the action was shown to reproduce many well-known equations of motion, e.g. the Josephson-Anderson equation, the Euler equation, the London equation, the equation of superfluid fluxoid quantization, and the d.c. Josephson equation. We have shown that the general current-phase relationship is consistent with the phase dynamics in a potential formally analogous to that of a \textit{rf} SQUID, consisting of quadratic, linear and sinusoidal terms whose relative strength is determined by the magnitudes of the charging and Josephson couplings, $ E_Q$ and $E_J$, respectively. The effective action leading to this current-phase relationship differs from that derived by AES in the context of superconductive tunneling~\cite{ambegaokar} due to the explicit presence of the parabolic potential in the action, as well as to the locality in time assumed in our analysis.  Analysis of dissipation in the aperture array deriving from the second order time nonlocal contribution of $G_{T}^{-1}$ will be addressed in future work.

The effect of the sinusoidal term in the current-phase relation was further analyzed using finite temperature renormalization group methods. We have shown that the sinusoidal part of the current-phase relationship is expected to become significant in two different regimes, but that it is most important when the coherence length $\xi(T)$ is larger than the characteristic size of the aperture, $\lambda_{ap}$. By exploiting the relationship between ${E_{Q} / E_{C}}$ and the ratio of the aperture size to the temperature-dependent healing length, we were able to examine the scaling of $E_J$ with respect to this ratio. This analysis identified regions II and III, separated by a singular line in the RG diagram, that are respectively consistent with the linear and sinusoidal current-phase relations that were observed experimentally in Ref.~\cite{packard}. Using the relevant experimental values of healing length and aperture dimensions, we have shown that the singular line separating these regions, $E_{Q}'=4\pi^{2}E_{C}'$, occurs about 20 mK below the lambda transition. The qualitative agreement of this value with the experimentally observed transition at $\sim$ 5 mK  below $T_{\lambda}$ in Ref.[\onlinecite{packard}] provides strong evidence for the validity of this effective theory. In addition, generalization of the effective action derived in this theory from one to two apertures shows that phase-difference coupling between multiple apertures leads to phase-difference synchronization and to a doubled amplitude of Josephson oscillation in the array.  Our analysis indicates that for $M$ parallel apertures in an array, we may expect the amplitude of Josephson oscillations to behave as $\mathcal{O}(M)$.

In this paper we have considered neither the dynamics of phase slips and the vortices by which they are carried, nor their role in the transition from linear to sinusoidal current-phase relationship (see Section \ref{subsubsection:velocity}).  However, we note that inclusion of the nonlocal interaction between $\omega(r)$ and $\omega(r')$ is expected to lead to the hydrodynamic equations first presented in reference [\onlinecite{khalatnikov}]. In the nanoaperture array, the low-temperature linear current-phase characteristic is thought to be due to independent nucleation and subsequent slippage of vortices at individual apertures~\cite{packard}. These events dissipate the kinetic energy of the hydrodynamic resonator slowly, as opposed to large scale coherent phase slips occurring at higher temperatures. In this regime, the diameters of vortex cores are nearly as large as the apertures themselves. This suggests a physical picture of vortex proliferation at the nanoaperture array leading to coherent oscillations. Such a picture is consistent with our requirements that i) $E_{J}$ be relevant in order to observe Josephson oscillations, and ii) $E_{Q}$ be small so that the gauge-invariant phase difference is not pinned to an integer multiple of 2$\pi$.  In the core of vortices pinned at the array, off-diagonal long range order is destroyed and $\Delta\gamma$ is allowed to fluctuate away from $2\pi \ell$. The dynamics of the vortices may be studied by deriving their effective theory using boson-vortex duality\cite{lee}.  Such a study would be useful both to confirm in the dual picture the features of the phase diagram derived here, and to investigate the properties of a vortex condensate in an aperture array for which the bosonic field operator used here no longer describes particles above the vacuum. 

In utilizing the current approach to interpret experiments on liquid $^{4}$He, we have neglected the strongly-interacting nature of superfluid helium, i.e., we do not consider a realistic two-body potential.  Realistic studies for Josephson effects in liquid helium driven through nanoscale aperture arrays may be undertaken with path-integral Monte Carlo methods~\cite{kwon_unpublished}. To our knowledge there has so far been no observation of Josephson oscillations between driven reservoirs of weakly-interacting condensed bosons separated by nanoaperture arrays, nor indeed of any Josephson effects under driving flow conditions for weakly interacting Bose condensate systems.  
However, the Josephson effect has been observed for weakly coupled Bose-Einstein condensates~\cite{oberthaler,levy}, and persistent currents have been observed in toroidally trapped condensates~\cite{ryu}. Taken together with the recently demonstrated ability to make arbitrary potentials in such geometries~\cite{boshier}, the rapid progress in experimental study and manipulation of rotating BECs in toroidal traps holds out the prospect of future realization of Josephson phenomena in confined atomic BECs.

\appendix
\section{\label{sec:pert} Perturbative expansion of $G^{-1}$}
For convenience and clarity, the perturbative expansion of $G^{-1}$ is included in this appendix. We use the following perturbative series to analyze the action Eq.~(\ref{eqn:greensaction}): {\small \begin{equation} \trlog{[G_{0}^{-1}+\delta G^{-1}]}=\trlog{[G_{0}^{-1}]} +\sum_{k=1}^{\infty}\frac{(-1)^{k+1}}{k}\tr{[(G_{0}\delta G^{-1})^{k}]} \label{eqn:expansion}\end{equation}}

The first term in this series is a constant which cancels due to the normalization of the partition function. The second term, a series in powers of $\delta G^{-1}$, gives important contributions to the effective action. The free Green's function of the action is found by inverting the $\tilde{G}_{0}^{-1}$ operator \cite{negele}  \begin{widetext}\begin{equation} \tilde{G}_{0}(k,\tau ;k',\tau ')=(2\pi)^{3}\delta(k-k')\exp [-\mathcal{E}_{k}(\tau -\tau ')](\Theta (\tau -\tau ')(1+n_{k})+ \Theta (\tau '-\tau )n_{k}) \nonumber \end{equation} \end{widetext}
where $\mathcal{E}_{k}=\frac{k^{2}}{2m}-\mu+V_{0}\Delta$ and $\displaystyle n_{k}=\frac{1}{e^{\beta\mathcal{E}_{k}}-1}$ is the Bose-Einstein distribution. When $\tau = \tau '$, the time-ordered correlation function is the normal ordered correlation function and so $\displaystyle \tilde{G}_{0}(k,\tau ;k',\tau )=n_{k}\delta(k-k')$.

In evaluating the integrations over internal momenta, we frequently use the fact that $\widetilde{G}_{0}(k,\tau ;k' ,\tau ')\propto \delta(k - k')$. Treating the perturbation series exactly results in nonlocal contributions to the action. We assume when needed that the imaginary time arguments of the higher order terms are the same, by appealing to the fact that the free Green's function is exponentially suppressed as distance in imaginary time increases. Momentum integrals over the free Green's function are restricted to $k=0$ because we are considering the low-energy dynamics of the condensed mode. 

\subsection{\label{subsection:thetaappen}$G_{\dot{\theta}}^{-1}$ contribution}

The $\trlog{}$ expansion with respect to this perturbation is outlined in the text (Section~\ref{subsection:Gtheta}). To convert Eq.(\ref{eqn:capacitive}) to a functional of $\Delta \gamma$ in the effective action, Eq.(\ref{eqn:capacitive}) is split (in position space) into left and right parts as $-{n^{2}V_{L}^{2} \over 4} \int_{0}^{\beta}d\tau \, \bigg( i\del_{\tau}\theta_{L} + m \phi_{L} \bigg)^{2} + (L\rightarrow R)$. Take $V_{L}=V_{R}\equiv V$ for simplicity. Using the Euler equation, $\phi_{L}$ is eliminated from the action. We then perform the Gaussian integral over $\phi_{R}$ to arrive at an effective term involving $\Delta \gamma(\tau)$ only. The Gaussian integral is: \begin{widetext} \begin{eqnarray}  \int \mathcal{D}[\phi_{R}]\exp \bigg[ -{1\over 2}\int_{0}^{\beta}d\tau \, V^{2}n^{2}m^{2}\phi_{R}^{2} - {Vn^{2}mi\over 2}\int_{0}^{\beta}d\tau \, \bigg( \del_{\tau}\theta_{L}+\del_{\tau}\theta_{R} + m\int_{r_{R}}^{r_{L}}dr\cdot \del_{\tau}v(r,\tau) \bigg) \phi_{R} \bigg] \nonumber \\ \propto  \exp \bigg[-{n^{2}V^{2} \over 8}\int_{0}^{\beta}d\tau \, \bigg( \del_{\tau}\theta_{L}+\del_{\tau}\theta_{R} + m\int_{r_{R}}^{r_{L}}dr\cdot \del_{\tau}v(r,\tau) \bigg)^{2} \bigg]  \end{eqnarray} \end{widetext} The $\phi_{R}$-independent part of the contribution is added into the exponent and the square expanded, yielding $S_{C}$ after simplification. Elimination of the 0-component of the gauge field from the action is reasonable because it is not a dynamical field.

\subsection{\label{sec:vappen}$G_{v_{g}}^{-1}$ contribution}

Analyzing the contribution of the vorticity energy density and the first order term in $G_{v_{g}}^{-1}$ to the perturbation expansion in Eq.~(\ref{eqn:expansion}) shows that the gauge-invariant velocity satisfies a London equation. We will use this equation to show that the first-order contribution of $G_{v_{g}}^{-1}$ to the effective action for $\Delta \gamma$ is canceled by the vorticity energy density term. Specifically, the first-order contribution to the action is 
\begin{eqnarray}
{} &{}& {1\over 2}\tr{\bigg[ \tilde{G}_{0}(k,\tau ;k,\tau )\cdot \frac{m}{2}\tilde{v}_{g}(q,\tau )\tilde{v}_{g}(-q,\tau ) \bigg] }\nonumber \\ &=&\frac{nm}{4}\int_{0}^{\beta}d\tau \int d^{3}r \;v_{g}(r,\tau)^{2}, \nonumber 
\end{eqnarray} where $n$ is the number of condensed bosons.  We have restricted the sum over momenta in the free Green's function to $k=0$ because contributions from $k\neq 0$ are exponentially suppressed at low temperatures. 

To derive the London equation Eq.~(\ref{eq:fairbank}) we set ${\delta S \over \delta v_{g}}=0$ at first order in the expansion in $G^{-1}$ and make use of the Euler equation, the identity $ \curl{(\curl{v_{g}(r,\tau)})}=\nabla(\nabla\cdot v_{g}(r,\tau))-\nabla^{2}v_{g}(r,\tau)$, and the physical requirement that $v(r,\tau)$ be divergence-free. Owczarek has exploited a similar ``Higgs"-type argument to rationalize the expulsion of circulation by a superfluid, noting that if the source of 
$v_{g}(r,\tau)$ is a roton, the penetration depth is roughly the same as experimentally-observed vortex core diameters.\cite{owczarek}

In Section \ref{subsubsection:velocity}, we stated that the first order contribution of $G_{v_{g}}^{-1}$ is canceled by the circulation energy density. This can be seen as follows: consider integrating the first order contribution along an integral curve $\Gamma$ of the superfluid velocity. Assuming a toroidal geometry as in the derivation of superfluid fluxoid quantization in Section~\ref{subsubsection:velocity}, we can approximate the first order contribution of $v_{g}$ by \begin{equation} {nm\over 4}\int d^{3}r\, v_{g}(r,\tau)^{2} \approx {nmL^{2}\over 4}\int_{\Gamma}ds\, v_{g}(r,\tau)^{2}, \end{equation} where the latter integral is with respect to arc-length and $L^{2}$ is the area factor multiplying the vorticity energy density in the microscopic Lagrangian. This integral can be converted to a line integral by identifying the tangent vector to $\Gamma$ with the driving velocity at each point. This is justifiable because i) $\Gamma$ is an integral curve of the superfluid velocity and it is physically reasonable to assume that for low $T$, $\nabla \theta (r,\tau)$ is parallel to $v(r,\tau)$ at each point in spacetime, and also since ii) $\Vert \nabla \theta(r,\tau) \Vert \ll \Vert v(r,\tau) \Vert$. For a constant magnitude driving velocity, the first order contribution becomes: \begin{eqnarray}\label{eqn:A5}  {nmL^{2}\over 4}\int_{\Gamma}ds\, v_{g}(r,\tau)^{2} &=&{mn\Vert v \Vert L^{2} \over 4} \int_{\Gamma}dr \cdot v_{g}(r,\tau)  \nonumber \\&=& {mn\Vert v \Vert L^{2} \over 4}\Phi_{0}(\ell -{\Delta\gamma(\tau) \over 2\pi}) \end{eqnarray}  
This contribution is canceled by the circulation energy density, which can be rewritten \begin{equation}{mL^{2}\over 2}\int d^{3}r \;v_{g}(r,\tau)j(r,\tau) \label{eqn:circdens} \end{equation} The vector identity $a \cdot (\curl{b})=b\cdot (\curl{a})-\nabla \cdot (a \times b)$ has been used in deriving this formula. Substituting into Eq.(\ref{eqn:circdens}) the London equation in the form $j(r,\tau)={-n\over 2L^{2}}v_{g}(r,\tau)$, one obtains the perturbation contribution in Eq.(\ref{eqn:A5}) but multiplied by a factor of $-1$. In using the vector identity above, we have neglected a topological contribution to the effective action. This is discussed in Section~\ref{subsubsection:velocity}.

\subsection{\label{sec:tunnappen}$G_{T}^{-1}$ contribution}

In this calculation, as in previous ones, we specialize to the left/right reservoir phase configuration $\theta(r,\tau) = \theta_{R/L}(\tau)$. In the multiaperture case, these become local left/right macroscopic phases in the vicinity of each aperture. Employing the convention that the left-to-right gauge-invariant phase difference is 
defined to be $-\Delta\gamma(\tau)$), the first order contribution is 
 \begin{eqnarray} {}&{}& {1\over 2}\tr{\bigg[G_{0}(r_{L},\tau; r_{R},\tau)T_{r_{R}r_{L}}e^{-i\Delta\gamma(\tau)} \bigg]} \nonumber \\ &+& {1\over 2}\tr{\bigg[G_{0}(r_{R}',\tau; r_{L}',\tau)T_{r_{L}'r_{R}'}e^{i\Delta\gamma(\tau)} \bigg]} \nonumber \end{eqnarray} The corresponding first-order contribution $S_{J}$ to the effective action results from using the fact that $\int d^{3}rd^{3}r' G_{0}(r,\tau ;r',\tau )=\tilde{G}_{0}(k-k'=0,\tau)$.

\begin{figure}
\begin{center}

\includegraphics[scale=.7]{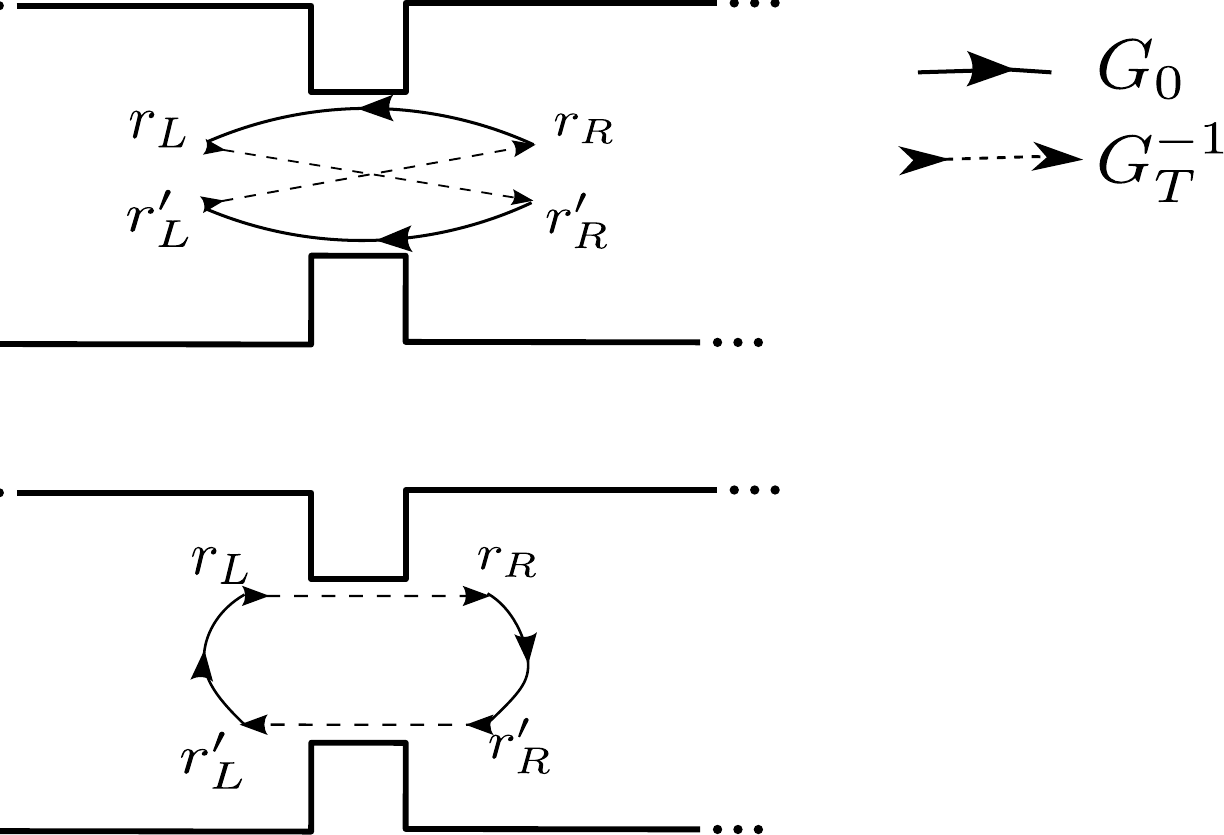}%
\caption{\label{second_tunnel} Diagrammatic representation of the two second-order contributions from $G_{T}^{-1}$ in Eq.(\ref{eqn:nonlocaltunn1}) which result in an imaginary time nonlocal contribution to the effective action. The convention $-\Delta \gamma$ is used for left-to-right hopping. We omit the delta function vertices as they are omitted in Eq.(\ref{eqn:nonlocaltunn1}).}
\end{center}  
\end{figure}

In our analysis of the effective theory for the gauge-invariant phase difference, nonlocal imaginary time terms in the perturbation expansion have been neglected. Here we derive one of these nonlocal terms arising from the second order contribution of $G_{T}^{-1}$; diagrams corresponding to this contribution are shown in Fig.~(\ref{second_tunnel}).

The resulting contribution is: \begin{widetext} \begin{eqnarray}\label{eqn:nonlocaltunn1} {}&{}& \exp{ \bigg[-{1\over 4}\bigg( \tr[ G_{0}(r_{L},\tau ; r_{R},\tau ')T_{r_{R}r_{L}'}e^{-i\Delta\gamma(\tau')} G_{0}(r_{L}',\tau '; r_{R}',\tau )T_{r_{R}'r_{L}}e^{-i\Delta\gamma(\tau)}  + (\mathrm{R}\leftrightarrow \mathrm{L})]  }\nonumber \\  &+& \tr[ G_{0}(r_{L},\tau ; r_{L}',\tau ')T_{r_{L}'r_{R}'}e^{i\Delta\gamma(\tau')}  G_{0}(r_{R}',\tau '; r_{R},\tau )T_{r_{R}r_{L}}e^{-i\Delta\gamma(\tau)}   + (\mathrm{R}\leftrightarrow \mathrm{L})] \bigg) \bigg] \end{eqnarray}\end{widetext}

Transforming to momentum space and taking the tunneling amplitude to be a constant, $T$, yields: \begin{eqnarray}\label{eqn:nonlocaltunn}
 -T^{2}\int_{0}^{\beta}d\tau \, d\tau ' \int d^{3}k\,d^{3}k' \, e^{({k'^{2}\over 2m}-{k^{2}\over 2m})(\tau -\tau ')}\\ (1+n_{k})n_{k'}\cos (\Delta \gamma(\tau)) \cos(\Delta \gamma (\tau'))\nonumber \end{eqnarray}

Our expression for the nonlocal contribution for this driven bosonic flow differs from that of AES because we do not have a particle-hole symmetry.  

\section{\label{sec:rgappen} RG for periodic potential}

We start by expressing the effective action 
Eq.~(\ref{eqn:effective}) in terms of Matsubara frequencies (we use the $S_{\mathrm{eff}}$ label for both the imaginary-time and Matsubara representations of the action): 
\begin{eqnarray} S_{\mathrm{eff}}[\widetilde{\Delta\gamma}; l, \beta] &= & 2\sum_{n=0}^{\infty}\bigg( E_{C}\omega_{n}^{2}-{E_{Q}\over 4\pi^{2}}\bigg)\widetilde{\Delta\gamma}(\omega_{n})\widetilde{\Delta\gamma}(\omega_{-n}) \nonumber \\ &+& \int_{0}^{\beta}d\tau \, E_{J}\cos \Delta\gamma(\tau) +{\beta E_{C}\ell \over 2\pi^{2}}\widetilde{\Delta
\gamma}(0). \label{eqn:matsu}\end{eqnarray}
Choosing a high-energy cutoff $\Lambda$, $\Delta \gamma (\tau)$ can then be split into low-frequency (slow, $s$) and high-frequency (fast, $f$) terms, $\Delta\gamma_{s}(\tau)=\int_{\vert \omega \vert < {\Lambda \over b}}{d\omega \over 2\pi } e^{-i\omega \tau}\widetilde{\Delta\gamma}(\omega)$ and $\Delta\gamma_{f}(\tau)=\int_{{\Lambda \over b}<\vert \omega \vert <\Lambda} {d\omega \over 2\pi} e^{-i\omega \tau}\widetilde{\Delta\gamma}(\omega)$, respectively, where $b$ is the renormalization scaling.  
The effective action is then split into slow ($s$), fast ($f$) and combination ($U$) components: 
\begin{eqnarray}S_{\mathrm{eff}}[\widetilde{\Delta\gamma}; l, \beta] &=&S_{s}[\widetilde{\Delta\gamma}]+S_{f}[\widetilde{\Delta\gamma}]+S_{U}[\Delta\gamma_{s}(\tau)+\Delta\gamma_{f}(\tau)] \nonumber \\
\mathrm{with}\,\, S_{U} [f(\tau)]&=& \int_{0}^{\beta}d\tau \, E_{J}\cos [f(\tau) ].
\end{eqnarray} 
We note that the slow part gets an additional contribution from the zero mode in Eq.~(\ref{eqn:matsu}).
Assuming that $T \ll \Lambda$, so that the Matsubara sums become integrals, it is then possible to 
integrate over the fast components by making use of a small $E_{J}$ approximation \cite{altland}
\begin{eqnarray} e^{-S_{\mathrm{low energy}}[{\Delta\gamma_{s}}]}&=&e^{-S_{s}[{\Delta\gamma_{s}}]}\langle 1- S_{U}[{\Delta\gamma_{s}},\Delta\gamma_{f}] + \ldots \rangle_{f}\nonumber \\ &\approx & e^{-S_{s}[{\Delta\gamma_{s}}]}e^{-\langle S_{U}[{\Delta\gamma_{s}},{\Delta\gamma_{f}}] \rangle_{f}},
\nonumber 
\end{eqnarray} 
to obtain an effective low energy action $S_{\mathrm{low energy}}[\Delta\gamma_{s}]$.  Here the ${f}$ subscript denotes an expectation value using $S_{f}$ as the action. 

An explicit evaluation of $\langle S_{U}[\Delta\gamma_{s},\Delta\gamma_{f}] \rangle_{f} $ results in a $b$-dependent multiplicative renormalization of $E_{J}$, which we call $E_{J}(b)$. This integration over fast modes is given explicitly by: \begin{widetext}\begin{eqnarray}
\langle S_{U}[{\Delta\gamma_{s}},\Delta\gamma_{f}]\rangle_{f} = E_{J}\int_{{\Lambda \over b} < \vert \omega \vert < \Lambda }\mathcal{D}[\widetilde{\Delta\gamma}(\omega)]e^{-2\int_{\Lambda \over b}^{\Lambda} {d\omega \over 2\pi}\bigg( E_{C}\omega^{2} -{E_{Q}\over 4\pi^{2}} \bigg) \vert \widetilde{\Delta\gamma}(\omega)\vert^{2}}\bigg( e^{i\Delta\gamma_{s}(\tau)}e^{i\int_{\Lambda \over b}^{\Lambda} {d\omega \over 2\pi} e^{i\omega \tau}\widetilde{\Delta\gamma}(\omega)-\mathrm{c.c.} }+\mathrm{c.c.} \bigg)
\end{eqnarray} 
\end{widetext}

Carrying out the Gaussian integration (and neglecting the divergent contributions) gives \begin{equation}\langle S_{U}\rangle_{f}= E_{J}\int_{0}^{\beta}d\tau e^{{1\over 4\pi^{2}}\int_{\Lambda \over b}^{\Lambda} {d\omega}{\pi \over E_{C}\omega^{2} - {E_{Q}\over 4\pi^{2}}}}\cos (\Delta\gamma_{s}(\tau)) \end{equation}

Rescaling $\tau \rightarrow {\tau \over b}$ to ensure that the Matsubara frequency still lies within the positive interval $[0,\infty )$ results in Eq.~(\ref{eqn:beta}) in the main text and we see that the periodic potential is multiplicatively renormalized. We have verified our result for the $\beta$-function using a background-field  RG analysis according to the procedure outlined in Ref.[\onlinecite{wen}].


\bibliography{iopreferences}

\begin{thebibliography}{10}
\bibitem{rgnote}%
  \BibitemOpen
  \bibinfo {note} {This result can also be obtained in a background field RG
  calculation by considering fluctuations around a low-energy $\Delta \gamma$
  configuration.}%
  \bibAnnoteFile{Stop}{rgnote}%
\end{thebibliography}

\end{document}